\documentclass[conference]{IEEEtran}
\makeatletter
\def\ps@headings{%
\def\@oddhead{\mbox{}\scriptsize\rightmark \hfil \thepage}%
\def\@evenhead{\scriptsize\thepage \hfil \leftmark\mbox{}}%
\def\@oddfoot{}%
\def\@evenfoot{}}
\makeatother
\ifCLASSINFOpdf
\usepackage[pdftex]{graphicx}
\else
\usepackage[dvips]{graphicx}
\fi

\usepackage[ruled,vlined]{algorithm2e}
\usepackage{hhline}
\usepackage{multirow}
\usepackage[normalem]{ulem}
\useunder{\uline}{\ul}{}
\usepackage{subcaption}

\usepackage{tikz}
\usepackage{pgfplots}
\usetikzlibrary{chains,decorations.text}
\usetikzlibrary{decorations.pathmorphing}
\usetikzlibrary{scopes}
\usepackage{cisco}

\usepackage{venndiagram}
\usepackage{amsmath}
\usepackage{url}
\usepackage{float}

\begin{document}
\title{Could Network View Inconsistency Affect Virtualized Network Security Functions?}
\author{
	\IEEEauthorblockN{Mohamed Aslan}
	\IEEEauthorblockA{Department of Systems and Computer Engineering\\ Carleton University. Ottawa, ON, Canada.\\
		maslan@sce.carleton.ca}
	\and
	\IEEEauthorblockN{Ashraf Matrawy}
	\IEEEauthorblockA{School of Information Technology\\ Carleton University. Ottawa, ON, Canada.\\
		ashraf.matrawy@carleton.ca}
}

\maketitle

\begin{abstract}
With SDN increasingly becoming an enabling technology for NFV in the cloud, many virtualized network functions need to monitor the network state in order to function properly.
An outdated network view at the controllers can impact the performance of those virtualized network functions.
In earlier work, we identified two main factors contributing to an outdated network view in the case of a load-balancer: network state collection and controllers' state distribution.
In this paper, we anticipate that the impact might be different in case of security functions. Therefore, we study the impact of an outdated network view on an anomaly-based IDS application.
In particular, we investigate: (1) the impact of controllers' state distribution on the performance of a distributed IDS in the case of a DDoS attack; and (2) the impact of network state collection on the performance of an IDS in the case of a TCP SYN flood attack.
Our results showed that the outdated network view had negative impact on the IDS anomaly-detection performance in the experiments that we conducted.
\end{abstract}


\IEEEpeerreviewmaketitle

\section{Introduction}
\IEEEPARstart{T}{he} Network Function Virtualization (NFV) is an emerging concept in cloud computing that enables network functions to run virtualized as software instances on the cloud.
Software-Defined Networking (SDN) is a promising network architecture, which simplifies the design of the virtualized network functions (VNF), and hence it is widely used in the cloud. In SDN, the control and data planes are decoupled. In order for applications to function properly, control-planes need to maintain a global view of the network.

Any inconsistency in the controller's network view might have an impact on applications' performance. In earlier work \cite{aslan2015impact}, we studied how the network state information collected by the controllers from the data-planes could negatively impact a load-balancer's performance. While, Levin \emph{et al.} \cite{levin2012logically} studied how the network state synchronization, in case of distributed controllers, among controllers could have a negative impact on a load-balancer's performance.
In this paper, we study the impact of network view inconsistency on a different network functions, an Intrusion Detection System (IDS).

In this paper, we make the following contributions:
(1) we investigate the impact of controllers' state distribution on the performance of a distributed anomaly-based IDS in the presence of a Distributed Denial of Service (DDoS) attack.
(2) we investigate the impact of network state collection on the performance of the IDS in the presence of a TCP SYN flood.
Our results showed that the IDS function that we evaluated was negatively impacted by the outdated information. Also, the type of the attack impacted how the outdated information affected the IDS. 

The rest of this paper is organized as follows:
In \S\ref{sec:related-work}, we provide a background on the topic and related work.
The setup of the IDS experiment is presented in \S\ref{sec:setup}.
We present two different attack scenarios and their results in \S\ref{sec:ddos} and \S\ref{sec:syn}.
Finally \S\ref{sec:conclusion} will be our conclusion and an outline for possible foreseeable work.

\section{Background and Related Work}\label{sec:related-work}

\subsection{Network State Collection Vs Controllers' State Distribution}
As aforesaid, SDN controllers need to maintain an up-to-date view of the network in order for the virtualized network functions to operate properly. To alleviate scalability and reliability issues in large-scale SDN deployments, recent work \cite{koponen2010onix,berde2014onos} encourages the use of physically distributed SDN controllers that are logically centralized. Therefore, controllers rely on two separate sources of state information: (1) data-planes (switches) which they control, and (2) other control-planes (controllers) in case of distributed deployments.

In case of single or distributed controller deployments, controllers need to collect network state information from the switches in a process known as \emph{network state collection}.
And in case of distributed controllers, those controllers also need to exchange network state information in order to keep their network view up-to-date, in a process known as \emph{controllers' state distribution}.

\subsection{Impact of Network View Inconsistency}
Together the network state collection and the controllers' state distribution processes could lead to inconsistency in the network state information at the controllers. Such inconsistency might have an impact on applications' performance.

In previous work \cite{aslan2015impact}, we showed that network state collection methods (\emph{active} or \emph{passive}) can have an impact on the performance of a load-balancer. Our results showed that the load-balancer's performance was affected by the \emph{polling} period of network state collection, in case of active network state collection. However, it was more resilient to the nature of the traffic load compared to passive \emph{flow}-based network state collection.

Using a distributed flow-based load-balancer, which employed a periodic synchronization of controllers' state information, Levin \emph{et al.} \cite{levin2012logically} demonstrated the negative impact of inconsistent network views at the controllers on the load-balancer's performance.
Two issues with Levin \emph{et al.}'s load-balancer \cite{levin2012logically} were identified by Guo et al. \cite{Guo201495}. First, short synchronization periods were required for acceptable load-balancing performance, leading to high synchronization overhead. Second, their design was vulnerable to forwarding loops due to the inconsistency of network views at the controllers.

In an attempt to mitigate the first issue with Levin \emph{et al.}'s load-balancer \cite{levin2012logically}, Guo et al. \cite{Guo201495} proposed a new trigger-based controller synchronization mechanism designed specifically for load-balancers called Load Variance-based Synchronization (LVS). LVS can perform well with lower synchronization overhead by \emph{only} synchronizing the controllers when the load of a specific server exceeds a certain threshold.

Subsequently, we can assume that the network state collection and the controllers' state distribution altogether could negatively impact the performance of virtualized network functions, more specifically load-balancers. Moreover, we anticipate that the impact might be different in case of different VNFs. Hence, in this paper, we study the impact of inconsistency on a different network function rather than a load-balancer. In particular, we study the impact of inconsistent network view on a security function, \emph{i.e.,} an IDS.

\section{Security Network Function}\label{sec:setup}
The IDS in our experiment is a network \emph{anomaly-based} IDS similar to that of \cite{munz2007traffic}, which was designed for non-SDNs. The IDS employs a clustering algorithm to learn about the behavior of network traffic and to detect the anomalous ones. Instead of using the NetFlow protocol and the classical K-Means clustering algorithm as in \cite{munz2007traffic}, it uses OpenFlow and employs a sequential K-Means clustering algorithm (shown in Algorithm \ref{algo:kmeans-seq}).
For active network state collection, our IDS implementation relies on OpenFlow STATS\_REQUEST messages.
In regards to controllers' state distribution, a simple delta-consistency model (\emph{i.e.,} periodic synchronization) was employed (see \cite{aslan2015impact} for more information).

To evaluate the impact of network view inconsistency on the IDS, we conducted two scenarios in order to simulate attacks on servers that are protected by the IDS. First, we used a set of the clients to launch a DDoS attack (\S\ref{sec:ddos}). Then, we randomly used one of the clients to launch a TCP SYN flood attack on the servers (\S\ref{sec:syn}). As for the traffic generation, Table \ref{ids-traffic-param} shows the parameters of the employed \emph{non-anomalous} traffic (attack traffic will be explained in \S\ref{sec:ddos}-A and \S\ref{sec:syn}-A).

\begin{algorithm}[h]
	\small
	\DontPrintSemicolon
	\SetKw{KwGoTo}{goto}
	\KwData{$P_{(p, b, f)}$, a point in the features vector space}
	\KwData{$T$, total number of points}
	\KwData{$M$, number of clusters}
	\KwData{$C_{i}$, the centroid of the $i^{th}$ cluster, $i \in [0, M)$}
	\KwData{$N_{i}$, number of points in the $i^{th}$ cluster}
	\SetKwFunction{nearest}{nearest}
	\Begin{
		$T$ $\leftarrow$ $\emptyset$\;
		\If {$T < M$} {
			$C_{T} \leftarrow P$\;
			$N_{T} \leftarrow N_{T} + 1$\;
		}
		\Else {
			$i_{c}$ $\leftarrow$ \nearest($P$, $C$)\;
			$C_{i_{c}} \leftarrow (C_{i_{c}} * N_{i_{c}}) + P$\;
			$N_{i_{c}} \leftarrow N_{i_{c}} + 1$\;
			$C_{i_{c}} \leftarrow \frac{C_{i_{c}}}{N_{i_{c}}}$\;
		}
		$T \leftarrow T + 1$\;
		
		\SetKwProg{nearest}{Function}{}{}
		\nearest{nearest (P, C)}{
			\KwData{$w_{k}$, weight of vector $k$, $k \in \{p, b, f\}$}
			\Begin{
				$idx \leftarrow \emptyset; min \leftarrow \infty$\;
				\For{$i \in [0, M)$}{
					$d \leftarrow \sqrt{(\frac{P_{p} - C_{i_{p}}}{w_{p}})^{2} + (\frac{P_{b} - C_{i_{b}}}{w_{b}})^{2} + (\frac{P_{f} - C_{i_{f}}}{w_{f}})^{2}}$\;
					\If{$d < min$}{
						$idx \leftarrow i$; $min \leftarrow d$\;
					}
				}
				\KwRet $idx$\;
			}
		}
	}
	\caption{Using Sequential K-means Clustering at the IDS.}
	\label{algo:kmeans-seq}
\end{algorithm}

\begin{table}[h]
	\caption{
		Employed legitimate traffic loads and their parameters.
		$r_{1}$ and $r_{2}$ are flow arrival rates for switch 1 and 2, respectively.
		$p_{1}$ and $p_{2}$ are messages-per-flow arrival rates for switch 1 and 2, respectively. Payload of any message is 512 bytes.
	}
	\centering
	\resizebox{.5\textwidth}{!}{
		\begin{tabular}{l | c}
			\multicolumn{1}{c}{} & Parameters\\
			\hhline{-:=:}
			Flows & \multicolumn{1}{c}{Poisson process}\\
			Flows rates &  $r_{1}=3 f/s, r{2}=4 f/s$\\
			Flow TTL & \multicolumn{1}{c}{2 sec}\\
			Msgs-per-flow & \multicolumn{1}{c}{Poisson process}\\
			Msgs-per-flow rates & $p_{1}=34 msg/s, p_{2}=30 msg/s$\\
			\hline
		\end{tabular}
	}
	\label{ids-traffic-param}
\end{table}

\section{DDoS Attack Scenario}\label{sec:ddos}
\subsection{Setup}
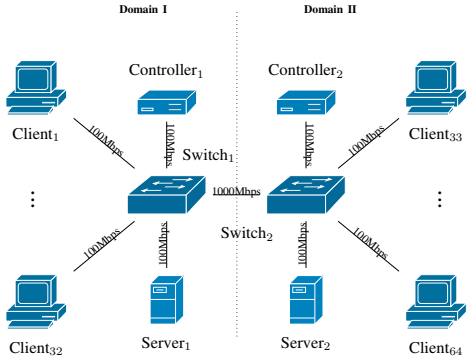
\begin{figure}[!t]
	\centering
	\resizebox{.35\textwidth}{!}{
		\begin{tikzpicture}[
		start chain=going right,
		diagram item/.style={
			on chain,
			join
		}
		]
		
		\node[label=above:Controller$_{1}$, diagram item, continue chain=going below] (controller1) {\controller};
		\node[label={[rotate=0]above right:$\kern-1em$Switch$_{1}$}, diagram item] (switch1) {\switch};
		\node[label=below:Server$_{1}$, start branch=1 going below, diagram item] (server1) {\server};
		\node[label=below:Client$_{1}$, start branch=2 going above left, diagram item] (client1) {\client};
		\node[label=below:Client$_{32}$, start branch=3 going below left, diagram item] (client32) {\client};
		\node[start branch=4 going left, on chain] (dots1) {\textbf{\vdots}$\qquad$};
		
		\node[label={[rotate=-0]below left:$\kern+1em$Switch$_{2}$}, diagram item, continue chain=going right] (switch2) {\switch};
		\node[label=above:Controller$_{2}$, start branch=5 going above, diagram item] (controller2) {\controller};
		\node[label=below:Server$_{2}$, start branch=6 going below, diagram item] (server2) {\server};
		\node[label=below:Client$_{33}$, start branch=7 going above right, diagram item] (client33) {\client};
		\node[label=below:Client$_{64}$, start branch=8 going below right, diagram item] (client64) {\client};
		\node[start branch=9 going right, on chain] (dots2) {$\qquad$\textbf{\vdots}};
		
		\scriptsize
		\draw[decoration={text along path, text={1000Mbps},text align={center}},decorate]  (switch1) -- (switch2);
		\draw[decoration={text along path, text={100Mbps},text align={center}},decorate]  (client1) -- (switch1);
		\draw[decoration={text along path, text={100Mbps},text align={center}},decorate]  (client32) -- (switch1);
		\draw[decoration={text along path, text={100Mbps},text align={center}},decorate]  (switch2) -- (client33);
		\draw[decoration={text along path, text={100Mbps},text align={center}},decorate]  (switch2) -- (client64);
		\draw[decoration={text along path, text={100Mbps},text align={center}},decorate]  (server1) -- (switch1);
		\draw[decoration={text along path, text={100Mbps},text align={center}},decorate]  (server2) -- (switch2);
		\draw[decoration={text along path, text={100Mbps},text align={center}},decorate]  (controller1) -- (switch1);
		\draw[decoration={text along path, text={100Mbps},text align={center}},decorate]  (controller2) -- (switch2);
		
		\draw [dotted] (1.5,-5) -- (1.5,2);
		\node at (-0.5,2) {\textbf{Domain I}};
		\node at (3.5,2) {\textbf{Domain II}};
		\end{tikzpicture}
	}
	\caption{The network topology with two distributed controllers.}
	\label{nettopo2}
\end{figure}

In this attack scenario, we used a number of the clients to simulate the behavior of a DDoS attack. We used a distributed (\emph{i.e.,} two controllers) \emph{active}-IDS implementation with a fixed polling period of $2$ seconds in this scenario. The network topology shown in Fig. \ref{nettopo2}. The IDS detects any anomalous behavior by monitoring the traffic arriving at the servers. We trained the IDS to expect traffic from an average of half the total number of clients at a given time. The features vector used for clustering is 3-tuple that is comprised of: (1) the total number of bytes received by a server ($b$), (2) the total number of packets received by a server ($p$), and (3) the total number of unique flows received by the server ($f$) per polling duration ($2$ sec). A weighted ($w_{b} = 4096$, $w_{p} = 64$, $w_{f} = 1$) Euclidean distance was used as a distance function for the clustering algorithm. We ran the experiment for 10 runs, in each run we simulated 5 DDoS (each attack lasts $30$ sec) attacks by allowing the dormant clients to send a CBR traffic to the servers.

\begin{figure}[!t]
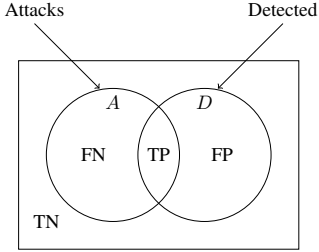

	\centering
	\resizebox{0.25\textwidth}{!}{
	\begin{venndiagram2sets}[labelA=$A$,labelB=$D$,labelOnlyA={FN},labelOnlyB={FP},labelNotAB={TN},labelAB={TP}]
		\setpostvennhook {
			\draw[<-] (labelA) -- ++ (135:2cm) node[above] {Attacks};
			\draw[<-] (labelB) -- ++ (45:2cm) node[above] {Detected};
		}
	\end{venndiagram2sets}
	}
	\caption{Venn Diagram for IDS performance indicators. TP is True Positives, TN is True Negatives, FP is False Positives, and FN is False Negatives.}
	\label{fig:ids-venn}
\end{figure}

In order to measure the performance of the IDS in this attack scenario, we use the following performance indicators:
(1) the \emph{time} to detect an anomalous behavior ($T_{\delta}$),
(2)
the \emph{precision} ($P_{\delta}$) (shown in (\ref{eqn:precision})),
the \emph{recall} ($R_{\delta}$) (shown in (\ref{eqn:recall})), and
the \emph{accuracy} ($A_{\delta}$) of detection (shown in (\ref{eqn:accuracy})), 
(3) and
the \emph{true positives} ($TP$),
the \emph{false positives} ($FP$), and
the \emph{false negatives} ($FN$) (shown in Fig. \ref{fig:ids-venn}).

\subsection{Results}
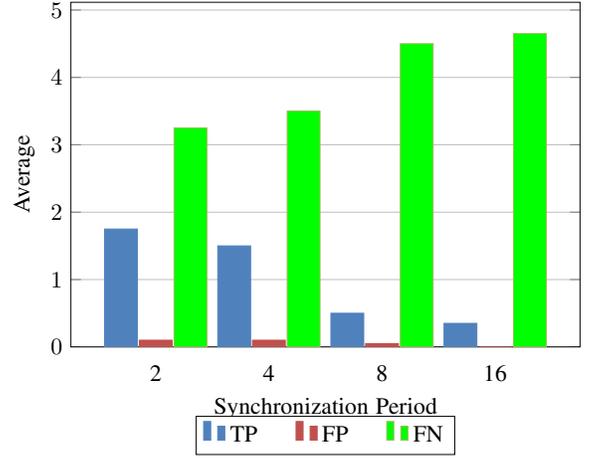
\begin{figure}[!t]
	\centering
	\resizebox{.45\textwidth}{!}{
		\definecolor{bblue}{HTML}{4F81BD}
		\definecolor{rred}{HTML}{C0504D}
		\definecolor{ggreen}{HTML}{9BBB59}
		\definecolor{ppurple}{HTML}{9F4C7C}
		
		\begin{tikzpicture}
		\begin{axis}[
		compat=1.3,
		width  = 0.5\textwidth,
		height = 6.75cm,
		major x tick style = transparent,
		ybar=2*\pgflinewidth,
		bar width=14pt,
		ymajorgrids = true,
		ylabel = {Average},
		xlabel = {Synchronization Period},
		symbolic x coords={2,4,8,16},
		xtick = data,
		scaled y ticks = false,
		enlarge x limits=0.25,
		ymin=0,
		legend cell align=left,
		legend style={
			at={(0.5,-0.2)},
			anchor=north,
			legend columns=-1,
			/tikz/every even column/.append style={column sep=0.5cm}
		}
		]
		
		\addplot[style={bblue,fill=bblue,mark=none}]
		coordinates {(2, 1.75)(4, 1.5)(8, 0.5)(16, 0.35)};
		\addplot[style={rred,fill=rred,mark=none}]
		coordinates {(2, 0.1)(4, 0.1)(8, 0.05)(16, 0.0)};
		\addplot[style={ggreen,fill=green,mark=none}]
		coordinates {(2, 3.25)(4, 3.5)(8, 4.5)(16, 4.65)};

		\legend{TP, FP, FN}
		\end{axis}
		\end{tikzpicture}
	}
	\caption{Average True Positives ($TP$), False Positives ($FP$), and False Negatives ($FN$) for Remote Servers (best viewed in color).}
	\label{fig:IDS2}
\end{figure}

\begin{figure}[!t]
	\resizebox{0.45\textwidth}{!}{
		\definecolor{bblue}{HTML}{4F81BD}
		\definecolor{rred}{HTML}{C0504D}
		\definecolor{ggreen}{HTML}{9BBB59}
		\definecolor{ppurple}{HTML}{9F4C7C}
		
		\begin{tikzpicture}
		\begin{axis}[
		compat=1.3,
		width  = 0.5*\textwidth,
		height = 6.75cm,
		major x tick style = transparent,
		ybar=2*\pgflinewidth,
		bar width=14pt,
		ymajorgrids = true,
		ylabel = {Detection Performance},
		xlabel = {Synchronization Period},
		symbolic x coords={2,4,8,16},
		xtick = data,
		scaled y ticks = false,
		enlarge x limits=0.25,
		ymin=0,
		legend cell align=left,
		legend style={
			at={(0.5,-0.2)},
			anchor=north,
			legend columns=-1,
			/tikz/every even column/.append style={column sep=0.5cm}
		}
		]
		
		\addplot[style={bblue,fill=bblue,mark=none}]
		coordinates {(2, 0.965)(4, 0.925)(8, 0.95833333333333)(16, 1.0)};
		\addplot[style={rred,fill=rred,mark=none}]
		coordinates {(2, 0.35)(4, 0.3)(8, 0.16666666666667)(16, 0.35)};
		\addplot[style={ggreen,fill=green,mark=none}]
		coordinates {(2, 0.34166666666667)(4, 0.295)(8, 0.16388888888889)(16, 0.35)};

		\legend{Precision, Recall, Accuracy}
		\end{axis}
		\end{tikzpicture}
	}
	\caption{Average precision ($P_{\delta}$), recall ($R_{\delta}$) and accuracy ($A_{\delta}$) vs synchronization period (best viewed in color).}
	\label{fig:IDS1}
\end{figure}
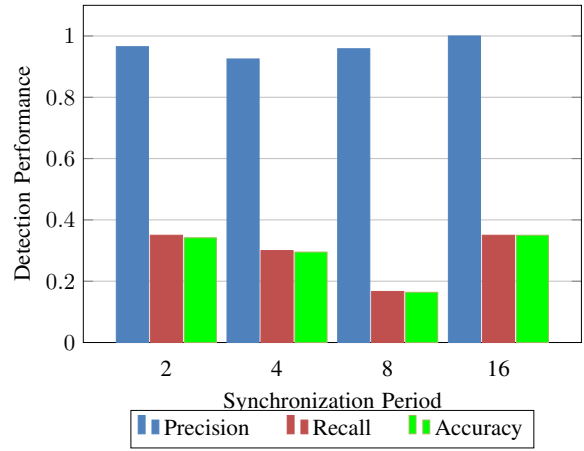

Fig. \ref{fig:IDS2} shows that as the synchronization period increases the true positives (TP) decreases while the false negatives (FN) increases indicating a degradation in the IDS performance. However, Fig. \ref{fig:IDS0} shows the synchronization period versus the average detection time ($T_{\delta}$). Fig. \ref{fig:IDS1} shows the effect of the synchronization period on the average precision ($P_{\delta}$), recall ($R_{\delta}$) and accuracy ($A_{\delta}$). The results shown is those figures do not necessarily show degradation in the IDS performance because we had to exclude those runs with \emph{infinite} detection time which increases as the synchronization period increases.  

\begin{align}
P_{\delta} = \frac{TP}{TP + FP} \label{eqn:precision}\\
R_{\delta} = \frac{TP}{TP + FN} \label{eqn:recall}\\
A_{\delta} = \frac{TP + TN}{TP + TN + FP + FN} \label{eqn:accuracy}
\end{align}

\begin{figure}[!t]
	\centering
	\resizebox{.4\textwidth}{!}{
		\begin{tikzpicture}
		\begin{axis}[xlabel=Synchronization Period (sec), ylabel=Average Detection Time (sec), xtick=data, grid=both, grid style=dotted,
		legend style={at={(0.5,-0.20)},anchor=north,legend columns=-1},
		symbolic x coords={2.0,4.0,8.0,16.0},height = 6.5cm]
		\addplot+[only marks, error bars/y dir=both, error bars/y explicit, mark=otimes*] coordinates {
			(2.0, 24.44791401724021) +- (0, 8.86664780675)
			(4.0, 12.958384533325832) +- (0, 5.64167164174)
			(8.0, 31.423265010118484) +- (0, 27.7394727404)
			(16.0, 17.774049091339112) +- (0, 8.59701241002)
		};
		\end{axis}
		\end{tikzpicture}
	}
	\caption{Average detection time ($T_{\delta}$) for Remote Servers.}
	\label{fig:IDS0}
\end{figure}
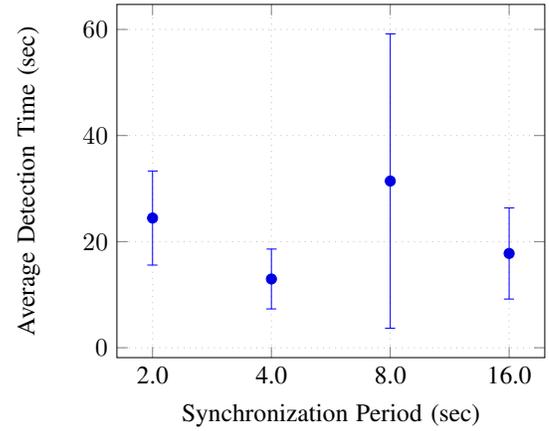

\section{TCP Syn Attack Scenario}\label{sec:syn}
\subsection{Setup}
In this attack scenario, we used one of the clients to launch a TCP SYN flood attack on the servers. We used a non-distributed (\emph{i.e.,} single controller) \emph{active}-IDS implementation in this scenario. The network topology shown in Fig. \ref{nettopo1}. The IDS detects any anomalous behavior by monitoring the traffic generated by the clients. We trained the IDS to expect traffic with parameters shown in Table \ref{ids-traffic-param}. The features vector used for clustering is 3-tuple that is comprised of: (1) the number of bytes per flow duration ($b$), (2) the number of packets per flow duration ($p$), and (3) the number of unique flows from the source ($f$).
A weighted ($w_{b} = 2046$, $w_{p} = 16$, $w_{f} = 0.25$) Euclidean distance was used as a distance function for the clustering algorithm. We ran the experiment for 10 runs, in each run we randomly select one of the clients to launch a TCP SYN flood attack.

\subsection{Results}
\begin{figure}[h]
	\centering
	\begin{tikzpicture}
	\begin{axis}[xlabel=Polling Period (sec), ylabel=Average Detection Time (sec), xtick=data, grid=both, grid style=dotted,
	legend style={at={(0.5,-0.20)},anchor=north,legend columns=-1},
	symbolic x coords={
		1.0,2.0,4.0,8.0,16.0,32.0,},height = 6cm]
	\addplot+[only marks, error bars/y dir=both, error bars/y explicit, mark=otimes*] coordinates {
		(2.0, 6.929869999999999) +- (0, 4.42041957281)
		(4.0, 9.351955) +- (0, 3.19380173528)
		(8.0, 14.9341) +- (0, 2.75844054332)
		(16.0, 19.418575) +- (0, 7.67982797621)
		(32.0, 39.24293) +- (0, 5.84604258918)
	};
	\end{axis}
	\end{tikzpicture}
	\caption{Average Detection Time vs Polling Period.}
	\label{fig:IDS3}
\end{figure}
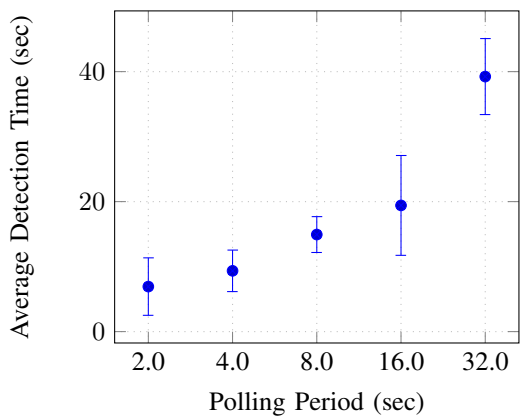

In order to measure the performance of the IDS in this attack scenario, we use the \emph{time} to detect an anomalous behavior ($T_{\delta}$) as our performance indicator.
Fig. \ref{fig:IDS3} shows the effect of the polling period on the performance of the IDS. The results show that as the polling period increases the time required to detect the anomalous behavior also increases. We believe that this is due the following reasons: (1) at the detection phase, the IDS fails to detect any attack that occurs between consequent polling periods, and (2) at the training phase, the polling period affects the number of collected data points (\emph{i.e,} the higher the polling period, the less the number of points collected) used during the IDS training phase, and hence affects the detection.

\begin{figure}[!t]
	\centering
	\resizebox{.4\textwidth}{!}{
		\begin{tikzpicture}[
		start chain=going right,
		diagram item/.style={
			on chain,
			join
		}
		]
		\node[label=above:Controller, diagram item] (controller) {\controller};
		{ [start branch=A going below left]
			\node[label={[rotate=0]below right:$\kern-1em$Switch$_{1}$}, diagram item] (switch1) {\switch};
			\node[label=below:Server$_{1}$, start branch=3 going below, diagram item] (server1) {\server};
			\node[label=below:Client$_{1}$, start branch=4 going above left, diagram item] (client1) {\client};
			\node[label=below:Client$_{32}$, start branch=5 going below left, diagram item] (client32) {\client};
			\node[start branch=6 going left, on chain] (dots1) {\textbf{\vdots}$\qquad$};
		}
		{ [start branch=B going below right]
			\node[label={[rotate=-0]below left:$\kern+1em$Switch$_{2}$}, diagram item, ,join=with switch1] (switch2) {\switch};
			\node[label=below:Server$_{2}$, start branch=7 going below, diagram item] (server2) {\server};
			\node[label=below:Client$_{33}$, start branch=8 going above right, diagram item] (client33) {\client};
			\node[label=below:Client$_{64}$, start branch=9 going below right, diagram item] (client64) {\client};
			\node[start branch=10 going right, on chain] (dots2) {$\qquad$\textbf{\vdots}};		
		}
		
		\scriptsize
		\draw[decoration={text along path, text={1000Mbps},text align={center}},decorate]  (switch1) -- (switch2);
		\draw[decoration={text along path, text={100Mbps},text align={center}},decorate]  (client1) -- (switch1);
		\draw[decoration={text along path, text={100Mbps},text align={center}},decorate]  (client32) -- (switch1);
		\draw[decoration={text along path, text={100Mbps},text align={center}},decorate]  (switch2) -- (client33);
		\draw[decoration={text along path, text={100Mbps},text align={center}},decorate]  (switch2) -- (client64);
		\draw[decoration={text along path, text={100Mbps},text align={center}},decorate]  (server1) -- (switch1);
		\draw[decoration={text along path, text={100Mbps},text align={center}},decorate]  (server2) -- (switch2);
		\draw[decoration={text along path, text={100Mbps},text align={center}},decorate]  (switch1) -- (controller);
		\draw[decoration={text along path, text={100Mbps},text align={center}},decorate]  (controller) -- (switch2);
		\end{tikzpicture}
	}
	\caption{The network topology with a single controller.}
	\label{nettopo1}
\end{figure}
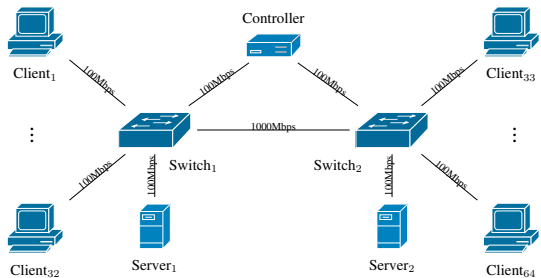

\section{Discussion and Conclusion}\label{sec:conclusion}
Previously \cite{aslan2015impact}, we articulated that load-balancers were affected by the network view inconsistency due to network state collection. In the case of TCP SYN flood attack (\S\ref{sec:syn}), the IDS exhibited the same behavior as the load-balacers \emph{i.e.,} the higher the polling period the worse the performance (due to reasons discussed in \S\ref{sec:syn}-B).
It is also worth mentioning that we tested the effect of synchronization period in the case of a distributed IDS for the TCP SYN flood attack scenario (\S\ref{sec:syn}). However, that effect was negligible compared to the effect of the polling period due to the \emph{localized} nature of the attack \emph{i.e.,} an IDS can still detect the presence of the attack even without any state synchronization between the controllers. Conversely, in the case of DDoS attack (\S\ref{sec:ddos}), the synchronization period impacted the IDS (like the load-balancer) due the nature of the attack \emph{i.e.,} malicious clients distributed their traffic through the two domains.

In summary, this paper demonstrated the impact of network view freshness at the SDN controllers on a security network function. Our results indicated that an out-of-date network view had negative impact on the security network function we evaluated. However, the impact differed with the attack scenario. For future work, we plan to investigate how this impact could be mitigated.

\section*{Acknowledgment}
The second author acknowledges support from the Natural Sciences and Engineering Research Council of Canada (NSERC) through the NSERC Discovery Grant program.

\bibliographystyle{IEEEtran}
\bibliography{references}

\begin{thebibliography}{1}
\providecommand{\url}[1]{#1}
\csname url@samestyle\endcsname
\providecommand{\newblock}{\relax}
\providecommand{\bibinfo}[2]{#2}
\providecommand{\BIBentrySTDinterwordspacing}{\spaceskip=0pt\relax}
\providecommand{\BIBentryALTinterwordstretchfactor}{4}
\providecommand{\BIBentryALTinterwordspacing}{\spaceskip=\fontdimen2\font plus
\BIBentryALTinterwordstretchfactor\fontdimen3\font minus
  \fontdimen4\font\relax}
\providecommand{\BIBforeignlanguage}[2]{{%
\expandafter\ifx\csname l@#1\endcsname\relax
\typeout{** WARNING: IEEEtran.bst: No hyphenation pattern has been}%
\typeout{** loaded for the language `#1'. Using the pattern for}%
\typeout{** the default language instead.}%
\else
\language=\csname l@#1\endcsname
\fi
#2}}
\providecommand{\BIBdecl}{\relax}
\BIBdecl

\bibitem{aslan2015impact}
M.~Aslan and A.~Matrawy, ``{On the Impact of Network State Collection on the
  Performance of SDN Applications},'' \emph{Communications Letters}, vol.~20,
  no.~1, pp. 5--8, 2015.

\bibitem{levin2012logically}
D.~Levin, A.~Wundsam, B.~Heller, N.~Handigol, and A.~Feldmann, ``Logically
  centralized?: state distribution trade-offs in software defined networks,''
  in \emph{Proc. of the first workshop on Hot topics in software defined
  networks}.\hskip 1em plus 0.5em minus 0.4em\relax ACM, 2012, pp. 1--6.

\bibitem{koponen2010onix}
T.~Koponen, M.~Casado, N.~Gude, J.~Stribling, L.~Poutievski, M.~Zhu,
  R.~Ramanathan, Y.~Iwata, H.~Inoue, T.~Hama \emph{et~al.}, ``Onix: A
  distributed control platform for large-scale production networks.'' in
  \emph{OSDI}, vol.~10, 2010, pp. 1--6.

\bibitem{berde2014onos}
P.~Berde, M.~Gerola, J.~Hart, Y.~Higuchi, M.~Kobayashi, T.~Koide, B.~Lantz,
  B.~O'Connor, P.~Radoslavov, W.~Snow \emph{et~al.}, ``Onos: towards an open,
  distributed sdn os,'' in \emph{Proceedings of the third workshop on Hot
  topics in software defined networking}.\hskip 1em plus 0.5em minus
  0.4em\relax ACM, 2014, pp. 1--6.

\bibitem{Guo201495}
Z.~Guo, M.~Su, Y.~Xu, Z.~Duan, L.~Wang, S.~Hui, and H.~J. Chao, ``Improving the
  performance of load balancing in software-defined networks through load
  variance-based synchronization,'' \emph{Computer Networks}, vol.~68, no.~0,
  pp. 95 -- 109, 2014, communications and Networking in the Cloud.

\bibitem{munz2007traffic}
G.~M{\"u}nz, S.~Li, and G.~Carle, ``Traffic anomaly detection using k-means
  clustering,'' in \emph{GI/ITG Workshop MMBnet}, 2007.

\end{thebibliography}

\end{document}